\documentclass[aps,amsmath,preprint]{revtex4}
\usepackage{graphics,epsf,epsfig}
\usepackage{bm}

\begin{document}

\title{Shear induced grain boundary motion for lamellar phases
in the weakly nonlinear regime}
\author{Zhi-Feng Huang and Jorge Vi\~nals}
\affiliation{
School of Computational Science and Information Technology,
Florida State University, Tallahassee, Florida 32306-4120
}
\date{\today}

\begin{abstract}

We study the effect of an externally imposed oscillatory shear on 
the motion of a grain boundary that separates differently oriented
domains of the lamellar phase of a diblock copolymer. A direct numerical
solution of the Swift-Hohenberg equation in shear flow is used for the
case of a transverse/parallel grain boundary in the limits of weak
nonlinearity and low shear frequency. We focus on the region of
parameters in which both transverse and parallel lamellae are linearly
stable. Shearing leads to excess free energy in the transverse region
relative to the parallel region, which is in turn dissipated by net motion of
the boundary toward the transverse region. The observed boundary motion is a
combination of rigid advection by the flow and order parameter
diffusion. The latter includes break up and reconnection of lamellae, as well
as a weak Eckhaus instability in the boundary region for sufficiently
large strain amplitude that leads to slow wavenumber readjustment.
The net average velocity is seen to increase with frequency and strain
amplitude, and can be obtained by a multiple scale expansion of the
governing equations.

\end{abstract}

\maketitle

\section{Introduction}
\label{sec_intro}

Below an order-disorder temperature, nanoscale phases of various
symmetries can be found in block copolymer melts
\cite{fredrickson96,larson99}. For example, for diblock 
copolymers that consist of two chemically distinct but covalently bonded 
monomers, six distinct phases that result from microphase separation have been
documented \cite{fredrickson96}. They differ by the symmetry of their
composition modulation. One of the most actively investigated 
microphases is the lamellar phase, which can be observed around 50-50 
composition (symmetric mixture). However, when melts are processed by thermal
quenching or solution-casting from a disordered phase, a macroscopic 
size sample usually exhibits polycrystalline configurations comprised 
of many locally ordered but randomly oriented domains (grains) and 
large amounts of topological defects, such as grain boundaries, 
dislocations, and disclinations. Such state of
partial ordering is undesirable for many applications, with significant 
affects on e.g., the optical and mechanical performance of the material. 
Therefore, the realization and control of long-range orientational
order of domains has been of great interest in both experimental
and theoretical studies.

Several methods have been used experimentally to achieve
microstructural alignment in block copolymer melts since the
discovery of flow-induced alignment by Keller \textit{et
al.} \cite{keller70}. One of the most 
common ways to induce global order in bulk samples is to 
impose an oscillatory or steady shear between two parallel plates, 
which not only has the advantage of easily characterizing the shear
stress and monitoring the aligning progress, but also can 
bring some interesting new physics \cite{fredrickson96,larson99,%
hadziioannou79,koppi92,winey93,wiesner97,chen98,hamley01}. For the 
case of lamellar morphology that we are interested in here, the two
most often observed alignments relative to the imposed shear are parallel 
(with the lamellar layers parallel to the shearing plane), and 
perpendicular (with the lamellar layers normal to the vorticity 
of the shear flow). Both alignments have been 
observed in nearly symmetric diblock copolymer 
systems such as poly(ethylenepropylene)-poly(ethylethylene) (PEP-PEE) 
\cite{koppi92} and polystyrene-polyisoprene (PS-PI) \cite{winey93,%
wiesner97}. Which of these two possible 
alignments is selected in a given experiment depends on 
temperature, and shear strain amplitude and frequency 
\cite{larson99,chen98,hamley01}. 
A third alignment direction, the so called transverse
orientation in which the lamellar normal is parallel to the shear 
direction, has been found to coexist with parallel orientation for 
entangled poly(styrene-b-ethylenepropylene) (S-EP) diblock copolymer 
at high frequency and in the strong segregation limit \cite{pinheiro96}. 
Other studies have focused on the kinetics of global alignment and
have shown that the ordering rate increases with shear frequency,
strain amplitude, and temperature \cite{chen98,gupta96}. Experiments
on PS-PI diblocks further show 
nonlinear effects of the strain amplitude on the alignment rate, and 
that the time scale for the development of alignment exhibits a power law 
dependence on the strain amplitude, with an exponent equal to $-3$ 
or $-5$ depending on the stage of alignment and in different 
frequency regimes \cite{gupta96}. Furthermore, many experiments have 
indicated that the motion of topological defects plays an important 
role in the global alignment of microdomains under shear, including 
the evolution of kink band defects and tilt boundaries \cite{winey98}, 
as well as the migration and annihilation of partial focal conic 
defects, boundaries, and tilt walls \cite{chen97,chen98}.

A coarse-grained description of a diblock copolymer melt has been
used, both in the weak segregation \cite{leibler80,%
ohta86,fredrickson87} and strong segregation \cite{ohta86} regimes,
to theoretically understand the effect of shear flow and the
mechanisms of morphology evolution 
\cite{cates89,fredrickson94,drolet99,chen02,doi96,ren01}. The analytic
studies of Cates and Milner \cite{cates89}, and Fredrickson 
\cite{fredrickson94} focused on the order-disorder transition from
the isotropic state to the lamellar phase and the related alignment 
dynamics for systems subjected to steady shear flow. Below the transition 
temperature, the stability of uniform
lamellar structures under an oscillatory shear flow has been addressed 
in both two-dimensional (2D) \cite{drolet99} and three-dimensional
(3D) \cite{chen02} cases. Bifurcation diagrams, including
secondary instabilities (Eckhaus and zigzag) were given. These analyses 
found that although all three types of lamellar orientations could be
linearly stable under specific conditions, the stability range of 
the perpendicular orientation is larger than that of the parallel
one, which in turn is larger than the region of stability of the
transverse orientation. The perpendicular alignment was also shown to
be the preferred orientation following the decay of unstable parallel or transverse
lamellae. In addition to
these analytic work, computer simulations have been performed to
investigate the dynamics of lamellar alignment in bulk samples under 
steady \cite{doi96} or oscillatory \cite{ren01} shear flow. The
effects of strain amplitude and shear frequency on the degree of sample 
alignment have been examined, as well as annihilation
processes of defects (such as dislocations and disclinations)
\cite{ren01}. Also, domain 
coarsening in 2D diblock copolymer melts has been addressed although in 
the absence of shear \cite{boyer01R,boyer02,shiwa02}.
Little is known about the effect of shear on coarsening 
of block copolymers.

Few of the previous theoretical studies discussed above focused 
on the detailed dynamics and quantitative properties of topological 
defects motion under shear, which are crucial for the understanding 
of alignment and coarsening. In this paper we 
study the detailed mechanisms of grain boundary 
motion under an oscillatory shear flow building upon stability results 
of uniform lamellar patterns \cite{drolet99,chen02}. 
We use a simplified 2D configuration which involves only two lamellar 
domains of parallel and transverse orientations respectively, separated 
by a grain boundary, and focus on shears of small amplitude (less than 
$50\%$) and low angular frequency. Compared to previous 
numerical simulations \cite{doi96,ren01,boyer01R,boyer02,shiwa02}, 
the aspect ratio (defined as the ratio between 
the lateral extent of the system and the lamellar wavelength) is
larger, so that important dynamic features associated with grain 
boundary motion, such as diffusive relaxation of lamellae, phase 
shift of boundary velocity, and wavenumber adjustment in the transverse 
region, can be quantitatively analyzed. 
These features were absent in earlier work on lamellar (roll)
systems without shear \cite{cross93,manneville83,tesauro87,boyer01}, 
and we argue below that they are important when the system is under
oscillatory shear. 

This paper is organized as follows: In Sec. \ref{sec_model} 
we introduce a dimensionless model equation based on the Swift-Hohenberg 
equation to describe the dynamic evolution of symmetric diblock
copolymer melts under shear, and describe the grain boundary 
configuration used. The numerical results including grain
boundary velocity and 
lamellar wavenumber are presented in Sec. \ref{sec_num}. 
We derive amplitude equations governing the system evolution in 
Sec. \ref{sec_ampl}, and compare the results with the direct solutions
of Sec. \ref{sec_num}. Finally, in  Sec. \ref{sec_concl}, we summarize
our results and discuss the physical origin of the phenomena observed.

\section{Model equation and grain boundary configuration}
\label{sec_model}

The system under consideration is a symmetric diblock copolymer
melt below the order-disorder transition temperature $T_{\rm ODT}$. 
For length scales larger than the microscopic monomer 
scale (i.e., at a mesoscopic level) and time scales long enough compared
to the molecular relaxation of the polymer chains, a coarse-grained
description of the block copolymer melt can be used, with an order 
parameter field $\psi({\bf r},t)$ representing the local density
difference of the two constituent monomers. In the weak segregation
limit, i.e., close to $T_{\rm ODT}$, a coarse-grained free energy
functional has been derived \cite{leibler80,ohta86,fredrickson87}:
\begin{equation}
F[\psi] = \int d{\bf r} \left \{ -\frac{\tau}{2} \psi^2 + \frac{u}{4}
\psi^4 + \frac{\xi}{2} \left [ \left ( \nabla^2 + {q_0^*}^2 \right ) 
\psi \right ] ^2 \right \},
\label{eq_F}
\end{equation}
where $\tau$ denotes a reduced temperature variable which is a 
measure of the distance from the order-disorder transition and is 
positive below $T_{\rm ODT}$, and $q_0^*=2\pi/\lambda_0^*$ is the 
wavenumber of the periodic lamellar structure. Under
the assumption that changes in the local composition field 
$\psi$ are driven by the free energy minimization and advection by the flow,
the order parameter $\psi({\mathbf r},t)$ obeys a 
time-dependent Ginzburg-Landau equation
\begin{equation}
\frac{\partial \psi}{\partial t} + {\bf v} \cdot {\bf \nabla} \psi
= -\Lambda \frac{\delta F}{\delta \psi}.
\label{eq_G-L}
\end{equation} 
Here ${\bf v}$ is the local velocity field, $\Lambda$ is an Onsager 
kinetic coefficient and can be written as $\Lambda=M {q_0^*}^2$ 
(with $M$ a mobility).

We introduce a length scale $1/q_0^*$, a time
scale $1/\Lambda \xi {q_0^*}^4$ (which is the characteristic
polymer relaxation time and $\simeq 1/D{q_0^*}^2$, with $D$ the
chain diffusivity of the copolymer), and an order parameter scale 
$(\xi {q_0^*}^4 /u)^{1/2}$. Given PEP-PEE-2 as an example, we have
$\lambda_0^* \sim 30$ nm, $D \sim 10^{-11}$ cm$^2$/s for temperatures close
to $T_{\rm ODT}$, and hence the length scale here is about 5 nm and
the time scale is about 0.03 s. Consequently, the rescaled composition 
field $\psi$ obeys a dimensionless Swift-Hohenberg equation 
\cite{swift77} with an advection term
\begin{equation}
\frac{\partial \psi }{\partial t} + {\bf v} \cdot {\bf{\nabla}} 
\psi= \epsilon \psi - (\nabla^2 + q_0^2)^2 \psi - \psi^3,
\label{eq_s-h}
\end{equation}
where $\epsilon = \tau/\xi {q_0^*}^4$ and we have $0 < \epsilon \ll 1$
in the weak segregation regime considered here. Also, $q_0=1$
although the symbol $q_0$ is retained in what follows for clarity of presentation.

\begin{figure}
\centerline{\epsfig{figure=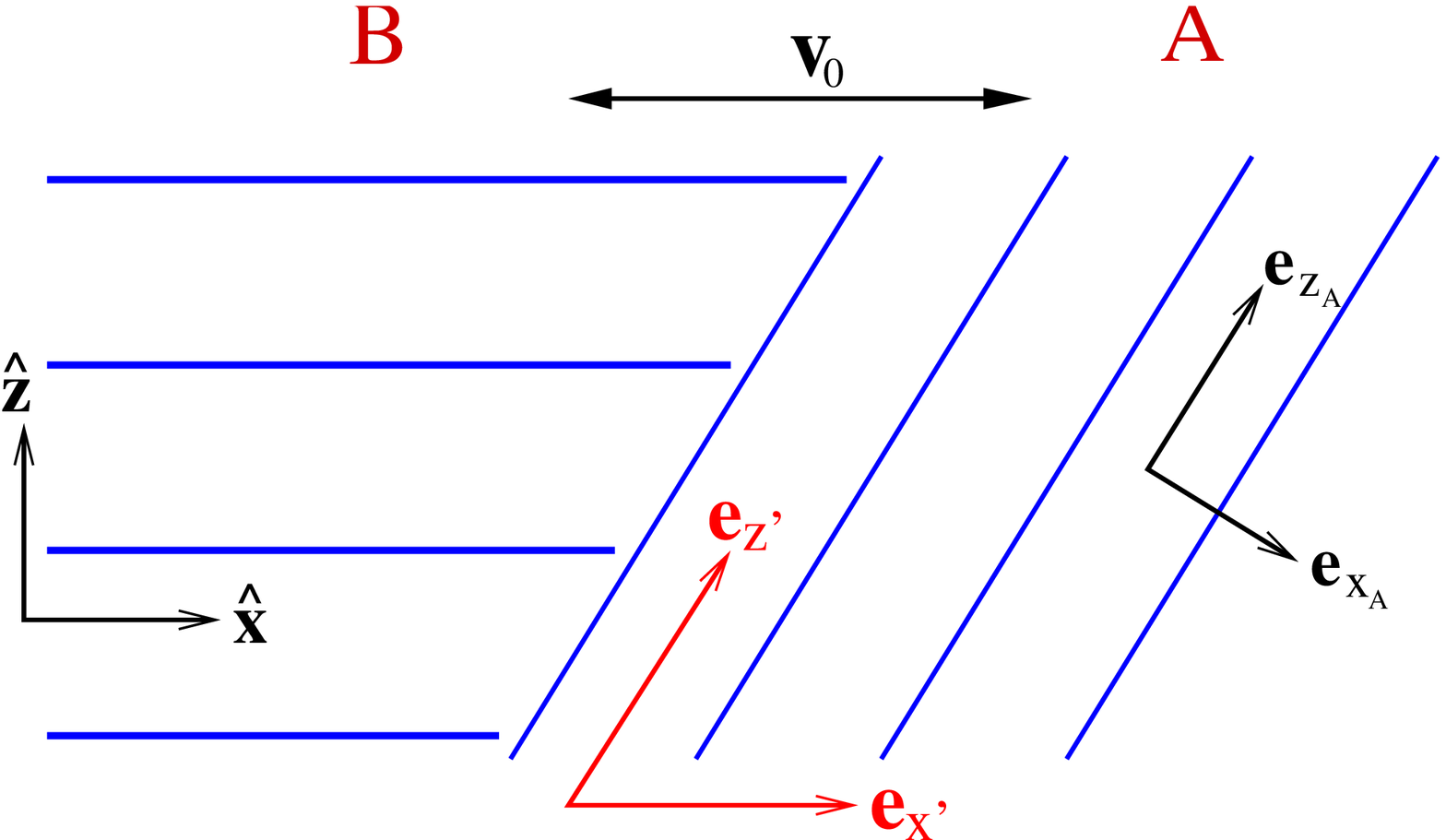,width=5in}}
\caption{Schematic representation of the two dimensional grain
boundary configuration under oscillatory shear flow studied in this
paper. Both the non orthogonal sheared frame $\{{\bf e}_{x'}, {\bf e}_{z'}\}$, 
and the auxiliary frame $\{{\bf e}_{x_{\rm A}}, 
{\bf e}_{z_{\rm A}}\}$ are indicated.
}
\label{fig_conf}
\end{figure}

As shown in Fig. \ref{fig_conf}, we 
consider a 2D reference state below the order-disorder transition 
containing a planar grain boundary that separates two semi-infinite 
ordered domains A and B. Initially both domains are in the 
lamellar state with the same wavenumber $q_0$ but 
oriented along different directions. We are interested here in the
case of a 
$90^{\circ}$ grain boundary with two mutually perpendicular lamellar 
sets A and B, a configuration that is known to be stable
against small perturbations in the absence of shear  
\cite{manneville83,tesauro87,cross93}. The two domains are under an
imposed shear flow
\begin{equation}
{\mathbf v}_0 = \frac{da}{dt} z {\bf \hat x}
=\gamma \omega \cos (\omega t) ~ z ~{\bf \hat x},
\label{eq_v_0}
\end{equation}
where $da/dt$ represents the shear rate with strain
$a(t)=\gamma \sin(\omega t)$, $\omega$ is the angular frequency, 
$\gamma$ the strain amplitude, and all quantities are assumed 
dimensionless. Lamellae of domain A are transverse (with the
wavevector components $q_x=q_0$ and
$q_z=0$) at $t=0$, and those in region B parallel ($q_x=0$ and 
$q_z=q_0$). Parallel lamellae B are marginal 
to the shear and not distorted, while transverse 
lamellae A are compressed, with both orientation and wavelength 
changing following the imposed shear as shown schematically in
Fig. \ref{fig_conf}. Thus, we anticipate that the 
grain boundary will not remain stationary even though both A 
and B lamellae are linearly stable under shear. Net motion results
from the free energy difference between region A (compressed) and 
region B (unchanged), as well as diffusive 
relaxation of the order parameter as shown below.

The stability of a uniform configuration of either parallel or
transverse lamellae under shear flow has been given in
Refs. \onlinecite{drolet99} and \onlinecite{chen02}. There exists a critical 
strain amplitude $\gamma_c$ above which the lamellar structure of a
given orientation melts, with $\gamma_c \rightarrow \infty$ for parallel
lamellae of wavenumber $q=q_0$, and small $\gamma_c$ for transverse
orientation. The stability diagrams presenting secondary instability
boundaries (for zigzag and Eckhaus modes) for 2D system have also
been given in Ref. \onlinecite{drolet99}. We focus below solely on shears 
for which both uniform parallel and transverse lamellae are linearly
stable. In addition, we consider the case in which shear effects are
of the same order of magnitude as diffusive relaxation of the order
parameter. Otherwise, at one extreme lamellae are passively advected
by the flow, whereas at the other, diffusion dominates.
If the velocity ${\bf v}$ in Eq. (\ref{eq_s-h}) can be approximated as ${\bf v}_0$
(which is the case under certain conditions \cite{fredrickson94},
including neglecting back flows due to osmotic stresses, and any
viscosity contrast between the microphases), then
the interesting range of $\gamma$ is such that the advection contribution
due to the imposed shear (${\bf v}\cdot {\bf \nabla} = (da/dt)z\partial_x$ 
in Eq. (\ref{eq_s-h})) is ${\cal O} (\epsilon)$. As will
be further discussed in Sec. \ref{sec_ampl} in connection with our
multiple scale analysis, this requires $\gamma \sim {\cal O}
(\epsilon^{1/4})$, an assumption that will be used in what follows.

\section{Numerical results}
\label{sec_num}

We first introduce a time dependent sheared frame of reference 
\cite{drolet99,chen02} in which the imposed shear flow vanishes. It is
defined by the
non-orthogonal basis set $\{ {\bf e}_{x'}={\bf \hat x}$, 
${\bf e}_{z'}=a(t){\bf \hat x}+{\bf \hat z} \}$ shown in Fig. 
\ref{fig_conf}. In this sheared frame, we have the coordinates 
$x'=x-a(t) z$ and $z'=z$. Also, the corresponding reciprocal basis 
set is defined as $\{ {\bf g}_{x'}={\bf \hat x}-a(t){\bf \hat z}$, 
${\bf g}_{z'}={\bf \hat z} \}$, with wavevector components expressed as 
$q_{x'}=q_x$ and $q_{z'}=a(t)q_x + q_z$. After coordinate 
transformation the Swift-Hohenberg equation (\ref{eq_s-h}) 
becomes
\begin{equation}
\frac{\partial \psi }{\partial t} = \epsilon \psi 
- ({\nabla'}^2 + q_0^2)^2 \psi - \psi^3,
\label{eq_s-h'}
\end{equation}
where the modified Laplacian operator takes the form
$$
{\nabla'}^2 = \left[ 1+a^2(t) \right] \partial^2_{x'} - 2 a(t) \partial_{x'}
\partial_{z'} + \partial^2_{z'}.
$$
The critical value $\gamma_{c}$ for neutral 
stability is independent of the shear frequency $\omega$ and given by
\begin{equation}
\gamma_c = \left ( \frac{-b + \sqrt{b^2 - 4dc}}{2d} \right )^{1/2},
\label{eq_gamma_c}
\end{equation}
with $b=q_{x'}^2 (2{\beta}^2+4q_{z'}^2-2q_0^2)/2$, $c={\beta}^4-2q_0{\beta}^2
+q_0^4 -\epsilon$, and $d=3q_{x'}^4/8$ (here ${\beta}^2=q_{x'}^2 +
q_{z'}^2$). The results for secondary instabilities were already 
presented in Ref. \onlinecite{drolet99}.

We numerically solve Eq. (\ref{eq_s-h'}) with periodic
boundary conditions (in the sheared frame) by using the pseudo-spectral
algorithm described in Ref. \onlinecite{cross94}, and also detailed
in the appendix of Ref. \onlinecite{drolet99}. The equation has been
discretized on a $1024 \times 1024$ square grid in the
sheared frame. In most of our calculations, we choose a mesh
spacing $\Delta x'=\Delta z'=\lambda_0/8$, corresponding to 8 grid
points per wavelength $\lambda_0$ ($=2\pi/q_0$) and so $\Delta q
=1/128$ for the wavenumber spacing. A second order, semi-implicit time stepping
algorithm is used, with a time step $\Delta t=0.1$. In order to
remain within the weak segregation regime we set $\epsilon=0.04$ and
the angular frequency $\omega$ chosen is ${\cal O} (\epsilon)$. Initial
conditions are obtained by numerical solution of 
Eq. (\ref{eq_s-h'}) without shear, i.e., $a(t)=0$, from a
starting configuration consisting of two symmetric grain boundaries
located at $x'=L_{x'}/4$ and $3 L_{x'}/4$ (with $L_{x'}=1024 \Delta x'$ 
the extent of the system in the $x'$ direction), separating a parallel domain B 
from two surrounding regions of transverse lamellae. This
configuration is allowed to evolve without shear until a stationary
solution is reached. This stationary solution is used as the initial
condition for the integration with $a(t) \neq 0$.

\subsection{Grain boundary motion due to shear and diffusive
  relaxation of the order parameter}
\label{subsec_motion}

\begin{figure}
\centerline{\epsfig{figure=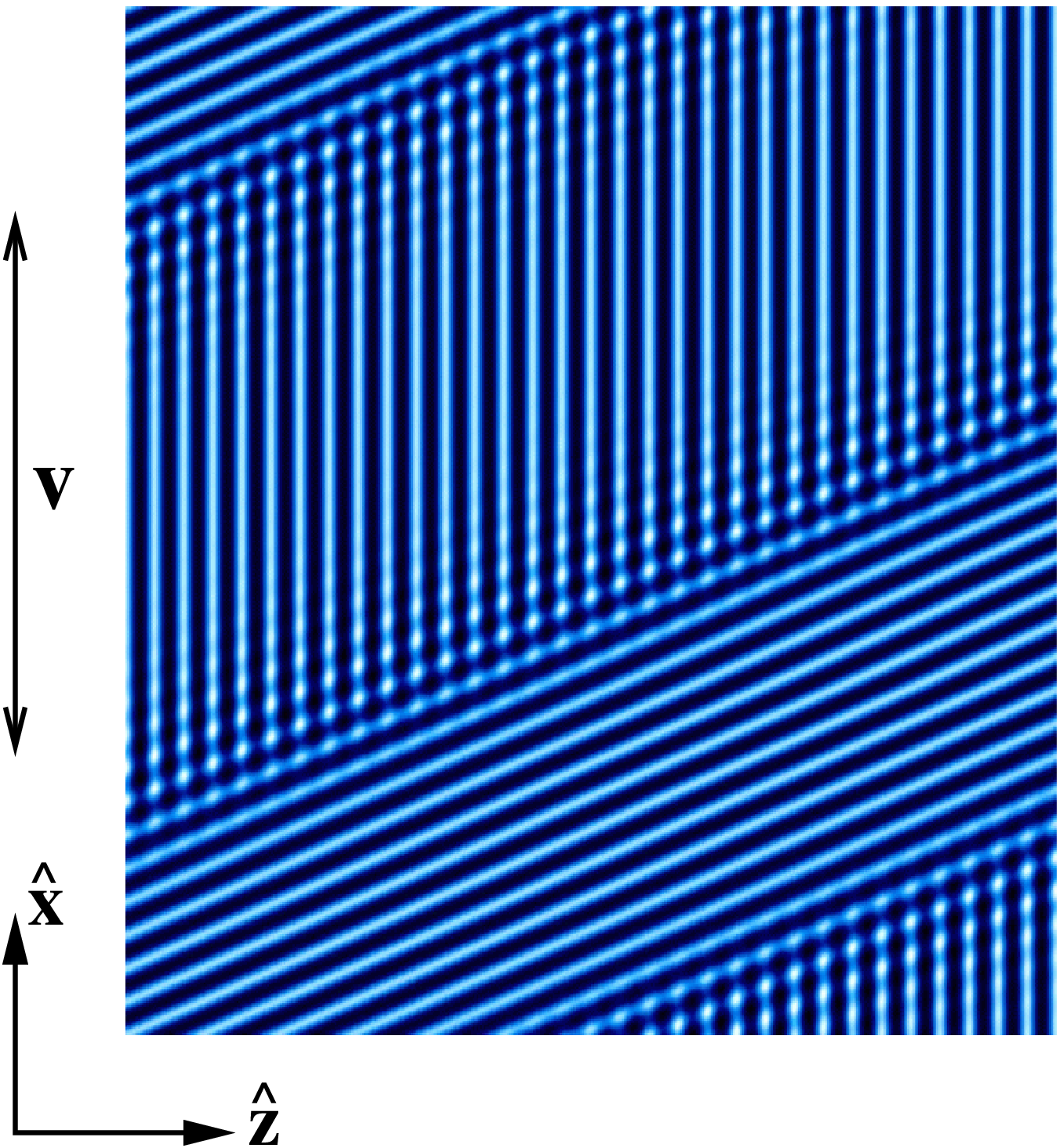,width=3in}}
\caption{Grain boundary configuration (in gray scale) at time $t=2.25 T_0$ 
obtained by numerically solving the Swift-Hohenberg equation
(\ref{eq_s-h'}) with $\epsilon=0.04$, 
$\gamma=0.4$, $\omega=0.04$, and $\Delta x = \lambda_0/32$. The 
configuration shown here has been transformed back to the laboratory frame 
$\{ {\bf \hat x}, {\bf \hat z} \}$. As time evolves, the transverse
domain is invaded by parallel lamellae, until a uniform parallel
configuration occupies the whole system.
}
\label{fig_shear}
\end{figure}

If $\gamma > \gamma_c= (8\epsilon/3)^{1/4}$, the stability limit of 
transverse lamellae (Eq. (\ref{eq_gamma_c})), A lamellae melt to 
a disordered state ($\psi=0$), and B lamellae
invade region A. If $\gamma < \gamma_{c}$, 
but still large enough to result in a secondary instability 
(Eckhaus, cross-roll, or zigzag) of transverse lamellae at a given 
frequency (e.g., $\gamma=0.5$ for $\epsilon=0.04$ and small $\omega$), 
our calculations show that small domains of parallel lamellae 
form within the bulk transverse region A as a result of the instability.
These domains evolve, and connect with each other and with
the approaching grain boundary to increase the extent of the parallel
region B. In both cases, the final configuration of the system is a uniform
lamellar structure of parallel orientation.
More interesting phenomena are observed in the 
range of $\gamma$ and $\omega$ in which both parallel and transverse
bulk regions are linearly stable, and when the contributions from
shear flow and order parameter diffusion are of the same order, as described
at the end of Sec. \ref{sec_model}. A typical transient configuration in this 
parameter range is shown in Fig. \ref{fig_shear}, with $\gamma=0.4$,
$\omega=0.04$, and grid spacing $\Delta x=\lambda_0/32$. The
configuration shown corresponds to $t=2.25 T_0$, where $T_0=2\pi / \omega$ 
is the shear period, and is presented in the laboratory frame basis set
$\{ {\bf \hat x}, {\bf \hat z} \}$. 

\begin{figure}
\centerline{\epsfig{figure=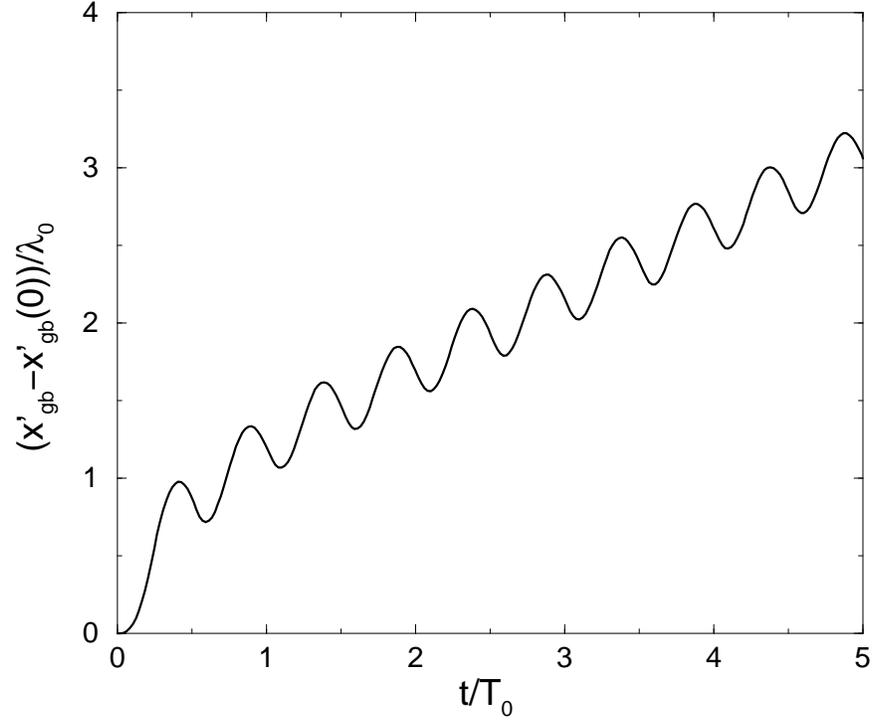,width=4.5in}}
\caption{Relative grain boundary displacement in the sheared frame 
$(x'_{\rm gb}-x'_{\rm gb}(t=0))/\lambda_{0}$ as a function of time $t/T_{0}$ for 
$\epsilon=0.04$, $\gamma=0.4$, $\omega=0.04$, and $\Delta x=
\lambda_0/8$.
}
\label{fig_xstar}
\end{figure}

Our analysis of the transient evolution of the configuration under
shear is based on the average location of the grain boundary $x'_{\rm
  gb}$. To determine this quantity we use a relation similar to that of Ref. 
\onlinecite{boyer01}: $B(x'_{\rm gb})=\sqrt{3} \sum_{i=1}^{n}
[\psi(x'_{\rm gb}, i\lambda_0) - \psi(x'_{\rm gb}, (i-1/2)\lambda_0)]
/4n = \delta$, with $n$ the number of pairs of lamellae in the $z'$ direction, 
and $\delta$ a quantity ${\cal O}(\epsilon)$. The value used here is
$\delta = \epsilon/4$. The grain boundary velocity $v'_{\rm gb}$ is defined as
the time rate of change of $x'_{\rm gb}$. Representative results 
(in the sheared reference frame) for $x'_{\rm gb}$ and $v'_{\rm gb}$
as a function of time are shown in Figs. \ref{fig_xstar} 
and \ref{fig_vgb}. Two distinct features can be clearly distinguished
as illustrated in Fig. \ref{fig_shear}:
First, transverse lamellae in the bulk rigidly follow the
oscillatory shear flow, leading to a periodic change 
in their orientation. Second, during part of the cycle the region of
parallel lamellae B grows as transverse lamellae in the boundary
region break up and reconnect as parallel lamellae (cross roll
instability). This process is partially reversed during the
rest of the cycle. The grain boundary exhibits oscillatory motion with
a nonzero net average as shown in Figs. \ref{fig_xstar} (location) and
\ref{fig_vgb} (velocity). Those portions of the shear cycle in which
broken transverse lamellae recombine correspond to the segments in
Fig. \ref{fig_xstar} with decreasing $x'_{\rm gb}$, or negative velocity
$v'_{\rm gb}$ in Fig. \ref{fig_vgb}. 

Whereas rigid distortion of transverse lamellae is the dominant response
to shear in the bulk, order parameter diffusion is important in the
boundary region, and is the mechanism that enables dissipation of
the excess free energy stored in the bulk transverse region due to shearing.
On the one hand, transverse lamellae are elastically compressed by the shear,
resulting in a net free energy increase in region A relative to B. As
a consequence, the elastic contribution to the system's energy is expected
to drive grain boundary motion toward the transverse domain
at all times during the shear cycle. On the other hand, backward
motion is also observed in Figs. \ref{fig_xstar} and \ref{fig_vgb} in
those portions of the shear cycle in which the magnitude 
of the shear strain $a(t)=\gamma \sin (\omega t)$ is small, indicating
diffusive relaxation of the order parameter field near the
grain boundary. The competition between the two determines the net
rate of advance of the boundary. This competition is illustrated in 
Fig. \ref{fig_vgb}(b). For $\gamma=0.3$ and
$0.4$ (with the same frequency $\omega=0.01$), the figure shows that
within one period $T_0$, a smaller strain amplitude $\gamma$ 
corresponds to larger temporal range of negative velocity, as well as
to a larger phase lag between the boundary velocity and the imposed
shear. The same effect can be seen in Fig. \ref{fig_vgb}(a), as lowering
the frequency enhances the effect of diffusion over elasticity.

The net velocity of the average location of the boundary is positive
as shown in Figs. \ref{fig_v_w} and \ref{fig_v_gamma}. The figures
plot the temporal average of velocity over a period $\langle
v'_{\rm gb} \rangle = \int_0^{T_0} dt 
v'_{\rm gb}(t) /T_0$ as a function of $\omega$ and $\gamma$.
The velocity increases sharply at very small $\omega$ and saturates at large
$\omega$. Diffusive relaxation is more pronounced at lower $\omega$,
consistent with the calculations shown in
Fig. \ref{fig_vgb}(a). $\langle v'_{\rm gb}\rangle$ also depends on strain amplitude
$\gamma$, as seen in Figs. \ref{fig_v_w} and \ref{fig_v_gamma}. The
average boundary velocity increases approximately as $\langle v'_{\rm gb}\rangle 
\sim \gamma^{\alpha}$ with $\alpha \sim 4$. The range of strain
amplitude accessible to our calculations is limited by the restriction that
$\gamma \sim {\cal O}(\epsilon^{1/4})$. As discussed above,
the transverse lamellar region is unstable for larger $\gamma$, whereas
for smaller $\gamma$, diffusion of order parameter dominates.
Note that the power law dependence is consistent with experiments
in PS-PI copolymers \cite{chen98,gupta96}, in which the 
rate of global alignment of a bulk sample is a power law of the strain
amplitude, with an exponent in the range $3-5$.

\subsection{Wavenumber adjustment in the transverse region}
\label{subsec_wavenumber}

Bulk transverse lamellae elastically compressed by the shear would have
a wavenumber $q_0 \sqrt{1+a^2(t)}$ in the laboratory frame, or 
$q_{x'}=q_0$, constant, in the sheared frame. Order parameter
diffusion in the boundary region, however, is seen to lead to a
wavenumber modification of transverse lamellae relative to what would
be expected from rigid deformation. We show in Fig. \ref{fig_qx}
the structure factor $|\psi_{q_{x'}}|$ along the $x'$ direction
defined as 
the Fourier transform of the order parameter $\psi$ at fixed $z'$ 
(e.g., at $z'=L_{z'}/2$, with $L_{z'}=1024 \Delta z'$ the system size 
along $z'$ direction). For large enough $\gamma$ ($\gamma > 0.1$), the peak
in $|\psi_{q_{x'}}|$ shifts away from $q_0=1$ with time,
asymptotically reaching a constant $q^m_{x'} < q_0$
(Fig. \ref{fig_qx}(a)). We have studied this wavenumber compression
for different frequencies $\omega$ and strain amplitudes $\gamma$. We
find that the 
value of $q^m_{x'}$ is independent of $\omega$, but that it increases
with decreasing $\gamma$ (Fig. \ref{fig_qx}(b)): 
$\delta q^m_{x'}=q_0-q^m_{x'} =5 \Delta q$ (with $\Delta
q=1/128$) for $\gamma=0.4$ (solid line), $ \delta q^m_{x'} = 3 \Delta q$ for
$\gamma=0.3$ (dotted line), $ \delta q^m_{x'} = \Delta q$ for
$\gamma=0.2$ (dashed line), and
$\delta q^m_{x'} = 0$ for $\gamma=0.1$ (thin solid line). These values
correspond to the disappearance of $5$, $3$, $1$, and $0$ 
lamellae in region A respectively. We further discuss this finding
in Sec. \ref{sec_ampl}, and in the discussion section 
\ref{sec_concl}. In addition to the motion in the position of the peak
of the structure factor, Fig. \ref{fig_qx}(a) also shows a decreasing
amplitude and broadening of the peak. This is due to the finite size
of the configuration; as the grain boundary moves, the region occupied
by transverse lamellae decreases.

The power spectrum of region B $|\psi_{q_{z'}}|$ as a function of
$q_{z'}$ (the Fourier
transform of $\psi$ at fixed $x'$ position, results not shown here)
is unaffected by the shear flow. Its maximum is located at
$q_z=q_{z'}=q_0$, without any visible shift in time.

\section{Multiple scale analysis and amplitude equations}
\label{sec_ampl}

A multiple scale analysis of the type frequently used to derive amplitude
equations close to instability thresholds
\cite{newell69,tesauro87,cross93} can be introduced here to derive an
equation of motion for the grain boundary in the limit of weak
segregation $\epsilon \ll 1$. As shown in Fig.
\ref{fig_conf}, we first define a time dependent, orthogonal 
basis set $\{ {\bf e}_{x_A} = ({\bf \hat x}-a(t){\bf \hat z})/
(1+a^2(t))$, ${\bf e}_{z_A} = (a(t){\bf \hat x}+{\bf \hat z})/
(1+a^2(t)) \}$ (different from the non orthogonal sheared frame 
$\{ {\bf e}_{x'}$, ${\bf e}_{z'} \}$ used in Sec. 
\ref{sec_num}). This frame is the orthogonal
frame of reference attached to the transverse region. The corresponding 
coordinates are $x_A=x-a(t)z$ and $z_A=a(t)x+z$. We then introduce an 
anisotropic coordinate scaling to define slowly varying amplitudes of 
$\exp(iq_0 x_A)$ as $X_A=\epsilon^{1/2} x_A$ and $Z_A=\epsilon^{1/4}
z_A$. We retain the laboratory frame coordinates in region B
$\{ {\bf \hat x}$, ${\bf \hat z} \}$ with a base mode given by
$\exp(iq_0 z)$. Its slowly varying amplitude is a function of the
rescaled variables $X_B=\epsilon^{1/4} x$ and $Z_B=\epsilon^{1/2} z$. 
We then expand the order parameter field $\psi$ as
\begin{equation}
\psi=\frac{1}{\sqrt{3}} \left ( A e^{iq_0 x_A} 
+ B e^{iq_0 z} +{\rm c.c.} \right ),
\label{eq_expan}
\end{equation}
where both complex amplitudes $A$ and $B$ are functions of the slow
spatial scales $X_A$, $Z_A$, $X_B$, and
$Z_B$, and of a slow time scale $T=\epsilon t$. The requirement of
slow amplitude change restricts our analysis to low frequencies $\omega
\sim {\cal O}(\epsilon)$, and we further focus on sufficiently small
shear amplitudes so that advection and local diffusion balance. This requires 
${\bf v} \cdot \nabla=(da/dt)z\partial_x \sim 
(\nabla^2+q_0^2)^2 \sim \epsilon$ according to Eq. 
(\ref{eq_s-h}). Considering that in the transverse region
$\partial_x=\partial_{x_A} + a \partial_{z_A}$, $\partial_z=
-a \partial_{x_A} + \partial_{z_A}$, $z=(-a x_A + z_A)/(1+a^2)$,
as well as the spatial $(x_A, z_A)$ and temporal scalings, we 
require that $\gamma \sim {\cal O}(\epsilon^{1/4})$.
The same relationship follows from the scalings appropriate for the
parallel region.

Following standard multiple scale procedure \cite{tesauro87,cross93}, 
we introduce the expansions $\partial_x \rightarrow \partial_{x_A}
+ \epsilon^{1/4} \partial_{X_B} + \epsilon^{1/2} (\partial_{X_A}
+ {\bar a} \partial_{Z_A})$, $\partial_z \rightarrow \partial_z
+ \epsilon^{1/4} (-{\bar a}\partial_{x_A} + \partial_{Z_A})
+\epsilon^{1/2} \partial_{Z_B} - \epsilon^{3/4} {\bar a} 
\partial_{X_A}$, and $\partial_t+(da/dt)z\partial_x \rightarrow
\epsilon [\partial_T + {\dot \gamma}_{\Omega} (X_A \partial_{Z_A}
+Z_B \partial_{X_B})] + {\cal O}(\epsilon^{5/4})$ (here ${\bar a}$
and ${\dot \gamma}_{\Omega}$ are defined by $a=\epsilon^{1/4}{\bar a}$
and $da/dt=\epsilon^{5/4}{\dot \gamma}_{\Omega}$), and we derive
the following amplitude equations at ${\cal O}(\epsilon^{3/2})$ from
Eq. (\ref{eq_s-h}),
\begin{eqnarray}
\partial_t A &=& \left [ \epsilon - \frac{da}{dt}
x_A \partial_{z_A} - \left ( 2iq_0 \partial_{x_A} + \partial_{z_A}^2 
- q_0^2 a^2 \right )^2 \right ]A - |A|^2 A -2|B|^2 A, 
\label{eq_A_local}\\
\partial_t B &=& \left [ \epsilon - \frac{da}{dt} 
z \partial_x - \left (\partial_x^2 + 2iq_0\partial_z \right )^2 
\right ] B -|B|^2 B - 2|A|^2 B.
\label{eq_B_local}
\end{eqnarray}
For $\gamma=0$, these equations reduce to those of Refs.
\onlinecite{manneville83} and \onlinecite{tesauro87}.

Equations (\ref{eq_A_local}) and (\ref{eq_B_local}) are expressed
in two different coordinates systems. We next transform them
to a common sheared frame $\{ {\bf e}_{x'}$, ${\bf e}_{z'} \}$,
by using the relations $x_A=x'$ and $z_A=ax'+(1+a^2)z'$. The base
state of the order parameter is still given by Eq. (\ref{eq_expan})
(with $x_A$ replaced by $x'$), and to ${\cal O}(\epsilon^{3/2})$, the resulting 2D
amplitude equations were already given in Ref. \onlinecite{huang03}.
Here we further assume a planar grain boundary in the sheared frame, 
for which the dependence of the amplitudes on the coordinate ($z'$) parallel 
to the grain boundary can be ignored. The complex amplitudes $A$ 
and $B$ satisfy the one-dimensional (1D) equations
\begin{equation}
\partial_t A = \left [\epsilon- ( 2 i q_0 \partial_{x'} 
-q_0^2 a^2 )^2 \right ] A -|A|^2 A -2 |B|^2 A, 
\label{eq_A}
\end{equation}
and
\begin{equation}
\partial_t B = \left [\epsilon- ( -2 i q_0 a \partial_{x'}
+ \partial^2_{x'} )^2 \right ] B -|B|^2 B -2 |A|^2 B.
\label{eq_B}
\end{equation}
At ${\cal O}(\epsilon^{3/2})$, two contributions from the shear flow
remain in Eqs. (\ref{eq_A}) and (\ref{eq_B}). The first one 
involves the term  $-2 i q_0 a \partial_{x'}$ in Eq. (\ref{eq_B}) 
and the term $(q_0^2 a^2)(2 i q_0 \partial_{x'})$ obtained from
the expansion of Eq. (\ref{eq_A}), which is non
negligible only in the grain boundary region and leads to diffusive
relaxation of the order parameter. The second is 
$q_0^4 a^4 A$ in Eq. (\ref{eq_A}). This term is uniform in the
entire region A, and reflects the contribution from advection of
transverse lamellae by the flow.

We had analytically calculated the velocity
of the grain boundary from these amplitude equations
in Ref. \onlinecite{huang03} by assuming
that the amplitudes can be approximated by 
$A(x',t) \simeq A(x'-x'_{\rm gb}(t))$ and
$B(x',t) \simeq B(x'-x'_{\rm gb}(t))$. We found that the velocity is
proportional to the 
free energy difference between the transverse and parallel phases, in
agreement with previous studies in the absence of flow 
\cite{manneville83,boyer01,boyer02}. Also, the results gave the
correct order of magnitude of the
average velocity, but we noted quantitative discrepancies
\cite{huang03}. We argued that the adiabatic approximation for
$A(x',t)$ and $B(x',t)$ given cannot incorporate diffusive relaxation
of the order parameter in the
boundary region so that the calculation only yields an upper
bound to the net boundary velocity. Since we have argued in 
Sec. \ref{sec_num} that this
relaxation is important, we turn here to a numerical determination of
the boundary velocity.

We present next the results of the numerical solution of
Eqs. (\ref{eq_A}) and (\ref{eq_B}). We initially consider
a region of parallel lamellae B surrounded by two regions 
of transverse lamellae A, and use periodic boundary conditions in the
integration. Both amplitudes $A$ and $B$ are complex variables. 
A pseudo-spectral method is applied, with a Crank-Nicholson scheme 
used for the linear terms, and a second-order Adams-Bashford scheme 
used for the nonlinear terms. The 
instantaneous location of the grain boundary $x'_{\rm gb}(t)$ is
defined by the condition $|B(x'_{\rm gb})|=\epsilon/4$, and its
velocity $v'_{\rm gb}$ is found by taking the time derivative of
$x'_{\rm gb}(t)$. In order to compare with the 2D 
results of the original model shown in Sec. \ref{sec_num}, we set the
system size $L=1024$, the time step $\Delta t=0.1$, and the grid 
spacing $\Delta x'=\lambda_0/8$. As was the case there, the initial
condition for $A$ and $B$ is provided by the steady solution of the
amplitude equations in the absence of shear.
Our results for the boundary velocity for $\gamma=0.4$, $\omega=0.01$,
and $\omega = 0.04$ are shown as dotted curves in
Fig. \ref{fig_vgb}(a), both in good
agreement with the direct solution of original model equation (\ref{eq_s-h'})
(symbols in the figure). The time averaged velocity $\langle v'_{\rm
  gb} \rangle$ is shown by the dashed lines in Figs. \ref{fig_v_w} and 
\ref{fig_v_gamma}.

In order to further analyze the wavenumber readjustment process discussed in
Sec. \ref{subsec_wavenumber}, we show our results in terms of the phase
$\phi_A$ of the complex amplitude $A$. In the sheared frame we define
$A=|A| \exp(i\phi_A)$ and plot $\phi_A$ as a function of
grid index $i_{x'} = x'/\Delta x'$ in Fig. \ref{fig_phaseA}.
Near the boundary, the phase becomes linear in space: $\phi_A
\propto -\delta q \cdot x'$ (the linear behavior is clearer
for a larger system size, as seen by comparing Fig. \ref{fig_phaseA}(a) 
($L=1024$) with \ref{fig_phaseA}(b) ($L=4096$)), indicating a local
wavenumber change $q_{x'} \rightarrow q_0-\delta q$. This is also in
agreement with the direct solution of the original model as shown in 
Fig. \ref{fig_qx}. Note also that the region of linearity (right side of
dot-dashed line in Fig. \ref{fig_phaseA}) increases with time, indicating that 
the re-adjustment of the local wavelength of the transverse lamellae 
first occurs at the boundary, and then progressively propagates into 
the bulk. At late times (e.g., $t=40 T_0$ in Fig. \ref{fig_phaseA}(a)), 
the wavenumber change can be observed in the whole transverse domain, 
with $q_{x'}$ corresponding to the stationary peak position 
of the structure factor $|\psi_{q_{x'}}|$ presented in
Fig. \ref{fig_qx}(a) ($t \geq 40 T_0$ there). We find that 
$\delta q = \delta q^m_{x'}$, the wavenumber shift discussed in
Sec. \ref{sec_num}, by determining $\delta q$ from the slope of the
dotted line in Figs. \ref{fig_phaseA}(a)and (b).

\section{Discussion and conclusions}
\label{sec_concl}

\begin{figure}
\centerline{\epsfig{figure=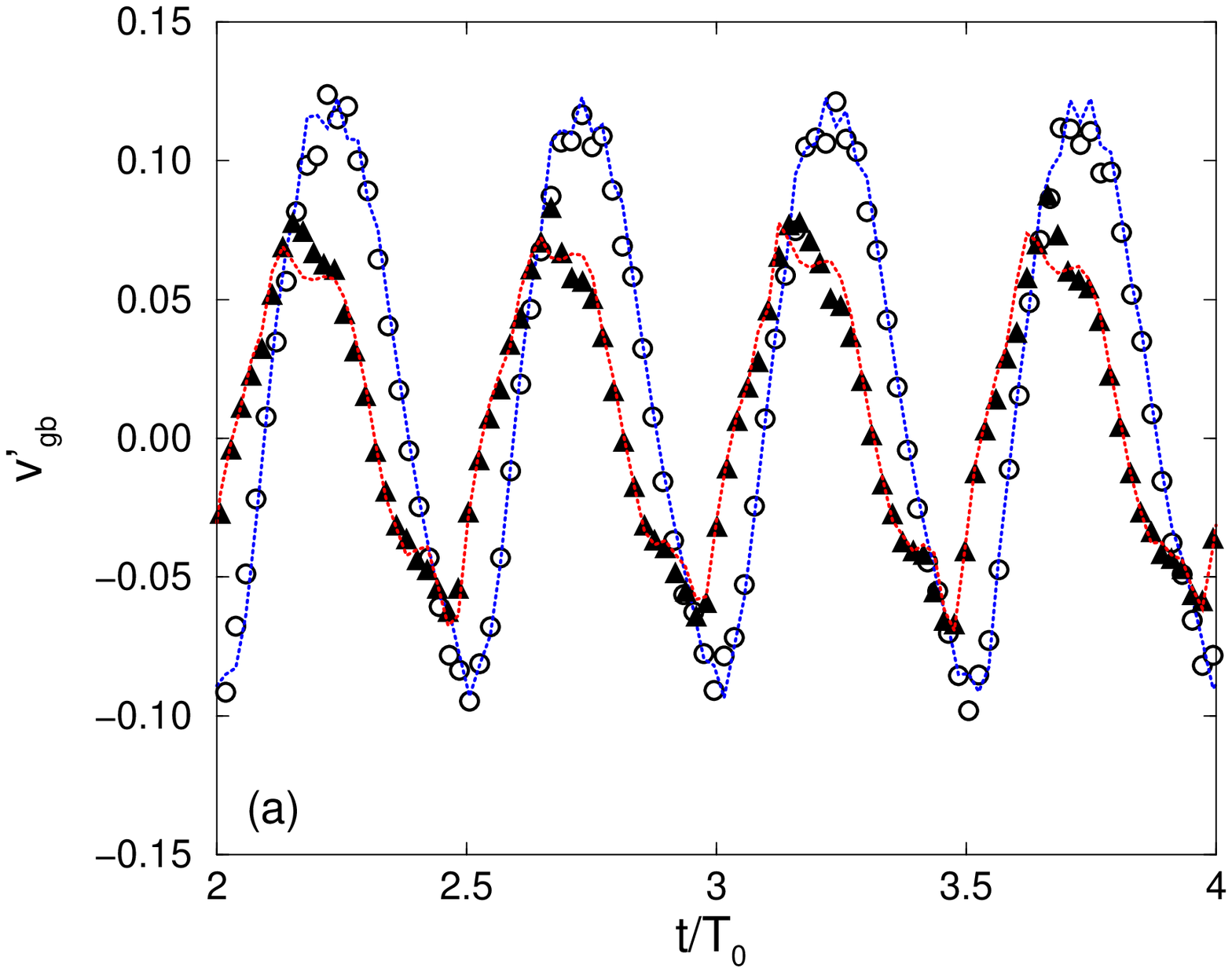,width=4.in}}
\centerline{\epsfig{figure=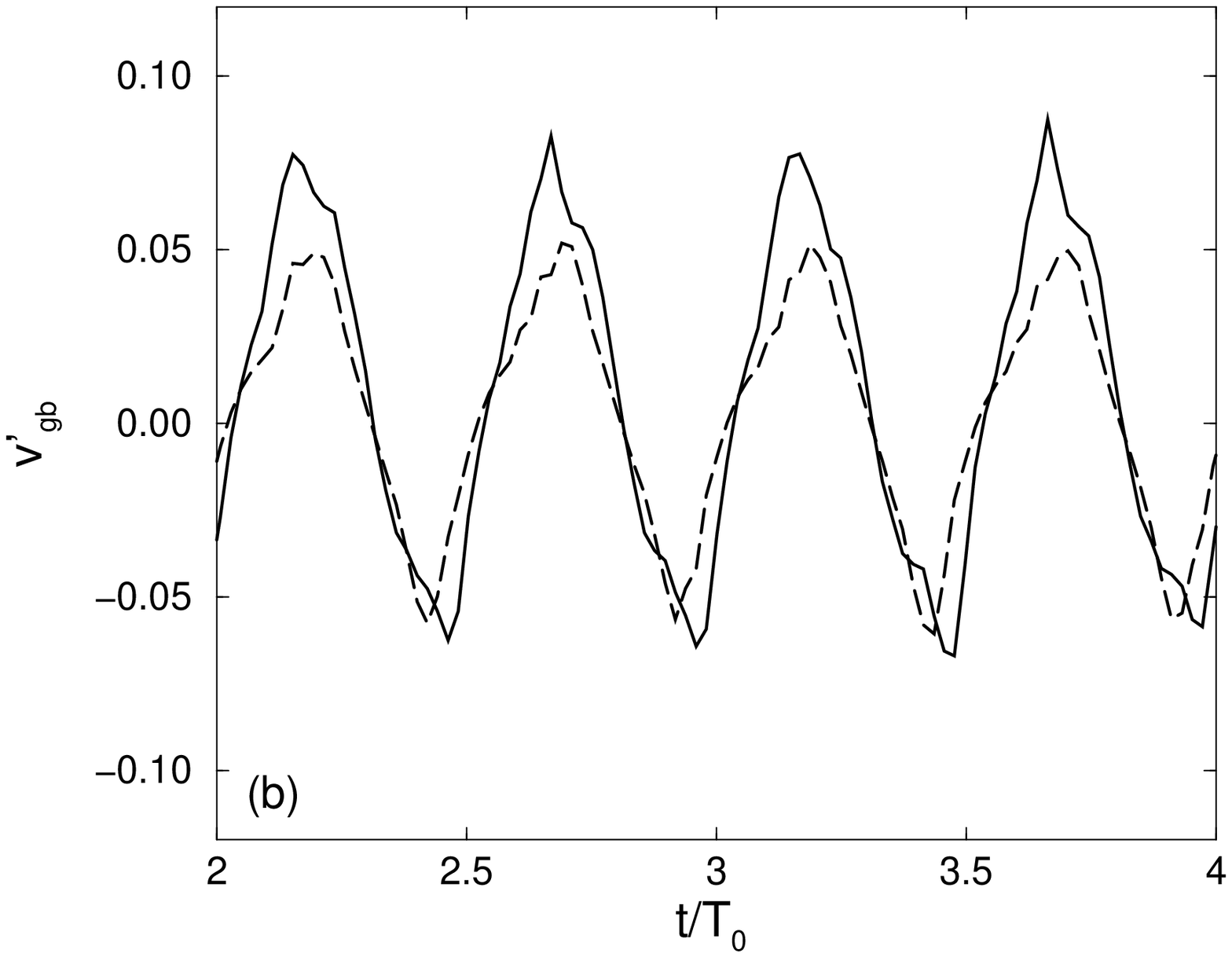,width=4.in}}
\caption{Boundary velocity $v'_{\rm gb}$ as
a function of time obtained from direct numerical integration of
Eq. (\ref{eq_s-h'}) with $\epsilon=0.04$ and different values of
$\omega$ and $\gamma$. (a) Fixed strain amplitude $\gamma=0.4$, 
$\omega=0.01$ (filled triangles), and $\omega = 0.04$ (open
circles). The dotted lines are the corresponding results obtained from
numerical integration of the amplitude equations (\ref{eq_A}) and (\ref{eq_B}).
(b) Results at constant frequency $\omega=0.01$, but different strain
amplitudes $\gamma=0.4$ (solid line), and $0.3$ (dashed line).
}
\label{fig_vgb}
\end{figure}

\begin{figure}
\centerline{\epsfig{figure=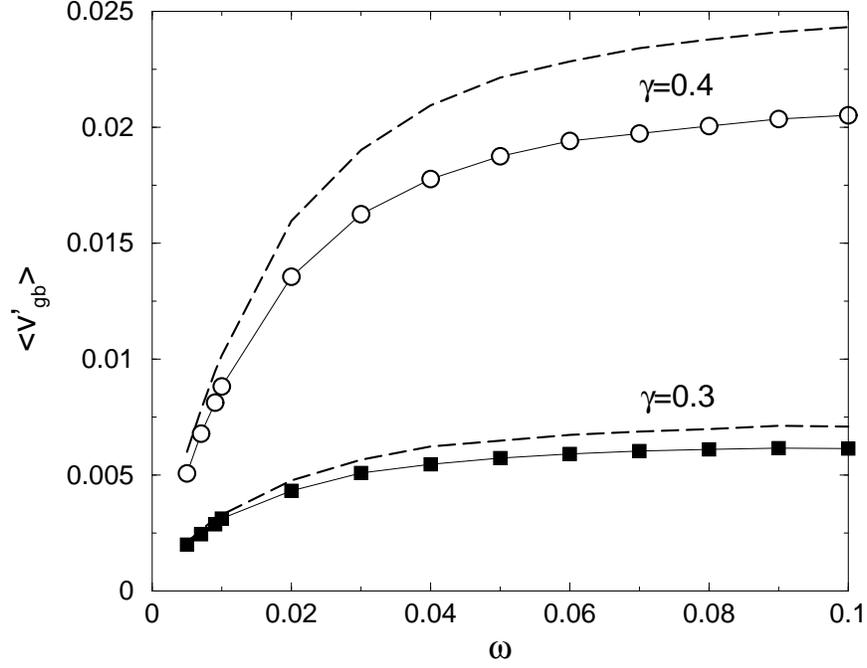,width=4.5in}}
\caption{Temporal average of the grain boundary velocity $\langle 
v'_{\rm gb}\rangle$ as a function of frequency $\omega$. The symbols
correspond to the solution of the order parameter model Eq. (\ref{eq_s-h'}) for
$\gamma=0.4$ (circles), and $0.3$ (squares), while the corresponding 
dashed lines are obtained from the amplitude equations (\ref{eq_A}) 
and (\ref{eq_B}).
}
\label{fig_v_w}
\end{figure}

\begin{figure}
\centerline{\epsfig{figure=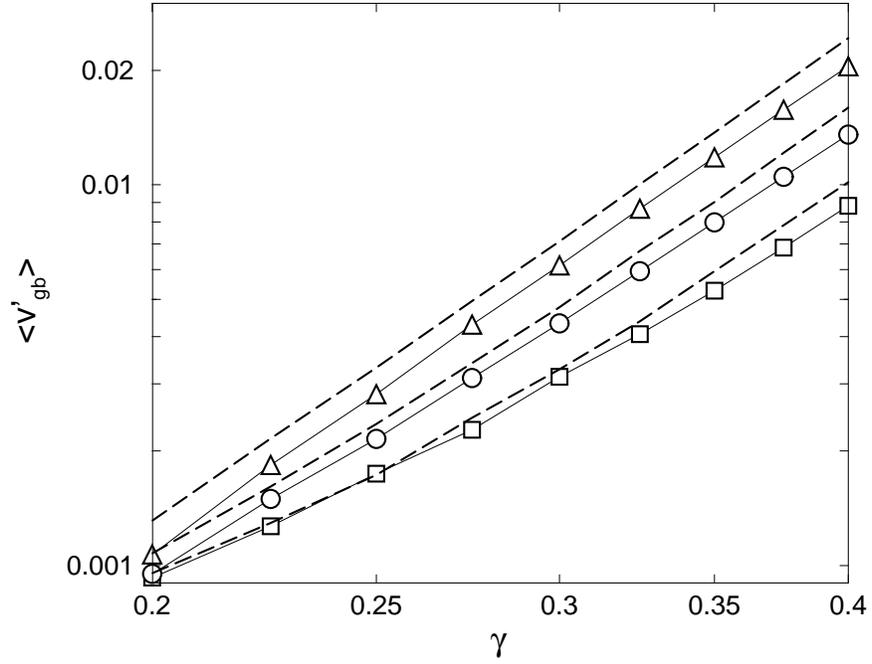,width=4.5in}}
\caption{Log-log plot of average velocity $\langle v'_{\rm gb}
\rangle$ versus strain amplitude $\gamma$ for $\omega=0.01$
(squares), $0.02$ (circles), and $0.1$ (triangles) from
direct solution of Eq. (\ref{eq_s-h'}). Also shown 
(dashed lines) are the corresponding results given by the amplitude
equations (\ref{eq_A}) and (\ref{eq_B}).
}
\label{fig_v_gamma}
\end{figure}

A coarse grained order parameter model has been used to study
the motion of a grain boundary separating two regions of uniform
parallel and transverse lamellae under an imposed shear flow. The
motion of the boundary is oscillatory, and
the driving force for motion is the excess energy stored in the elastically
strained transverse phase that can only be relieved through diffusive
relaxation of the order parameter in the boundary region. Diffusive
relaxation, however, is complex as the response of the order parameter
field is out of phase with the shear, and lamellae break up and reconnect
during each of the cycles. As expected, the effects of diffusive relaxation 
are more pronounced for small shear strain and low angular frequency,
as seen in both the time dependent behavior of boundary velocity (Fig.
\ref{fig_vgb}) or the time averaged velocity (Figs. \ref{fig_v_w} and
\ref{fig_v_gamma}). Although under the conditions of the study both
transverse and parallel orientations are linearly stable, we observe net
motion of the boundary toward the region occupied by transverse lamellae.

\begin{figure}
\centerline{\epsfig{figure=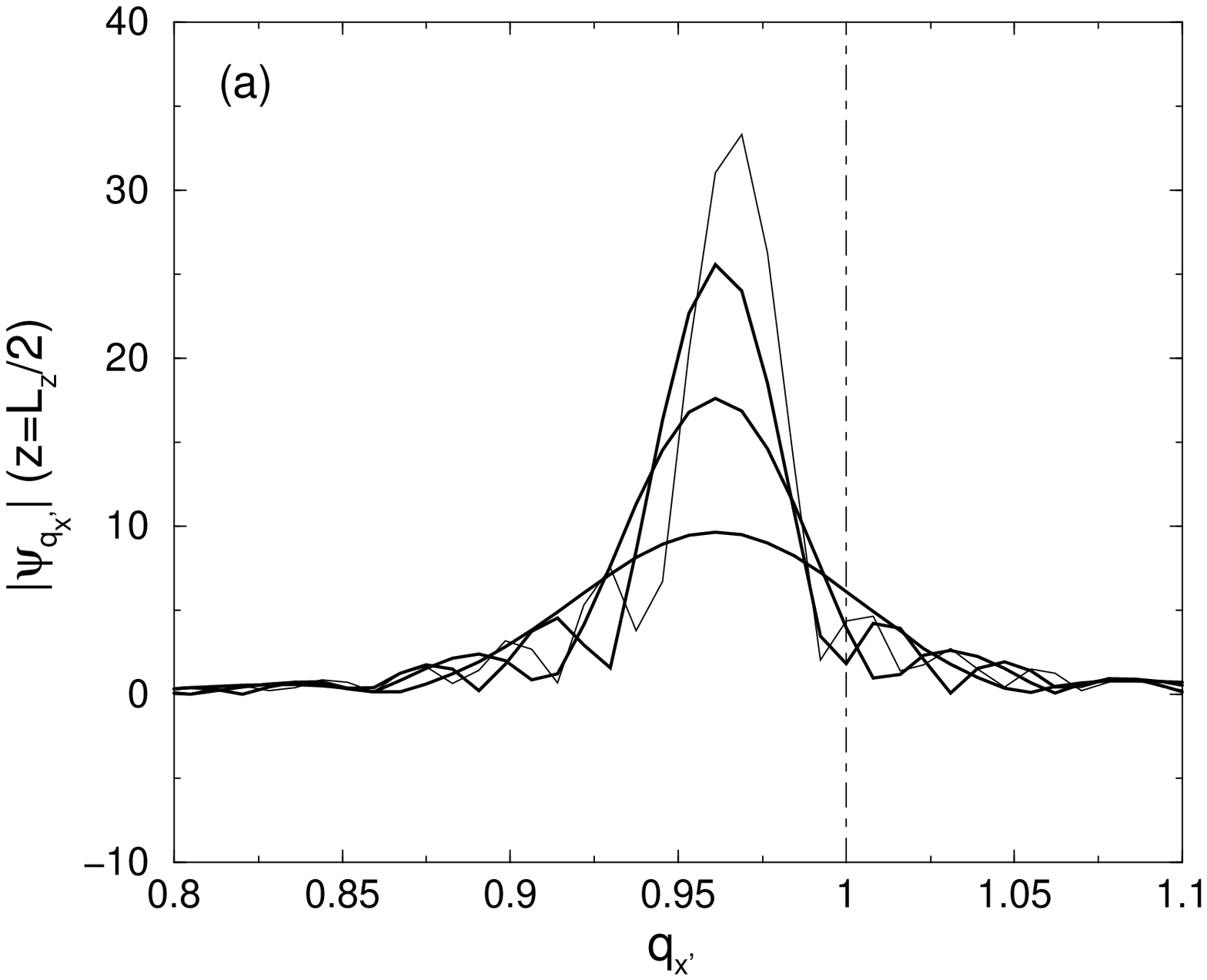,width=4.in}}
\centerline{\epsfig{figure=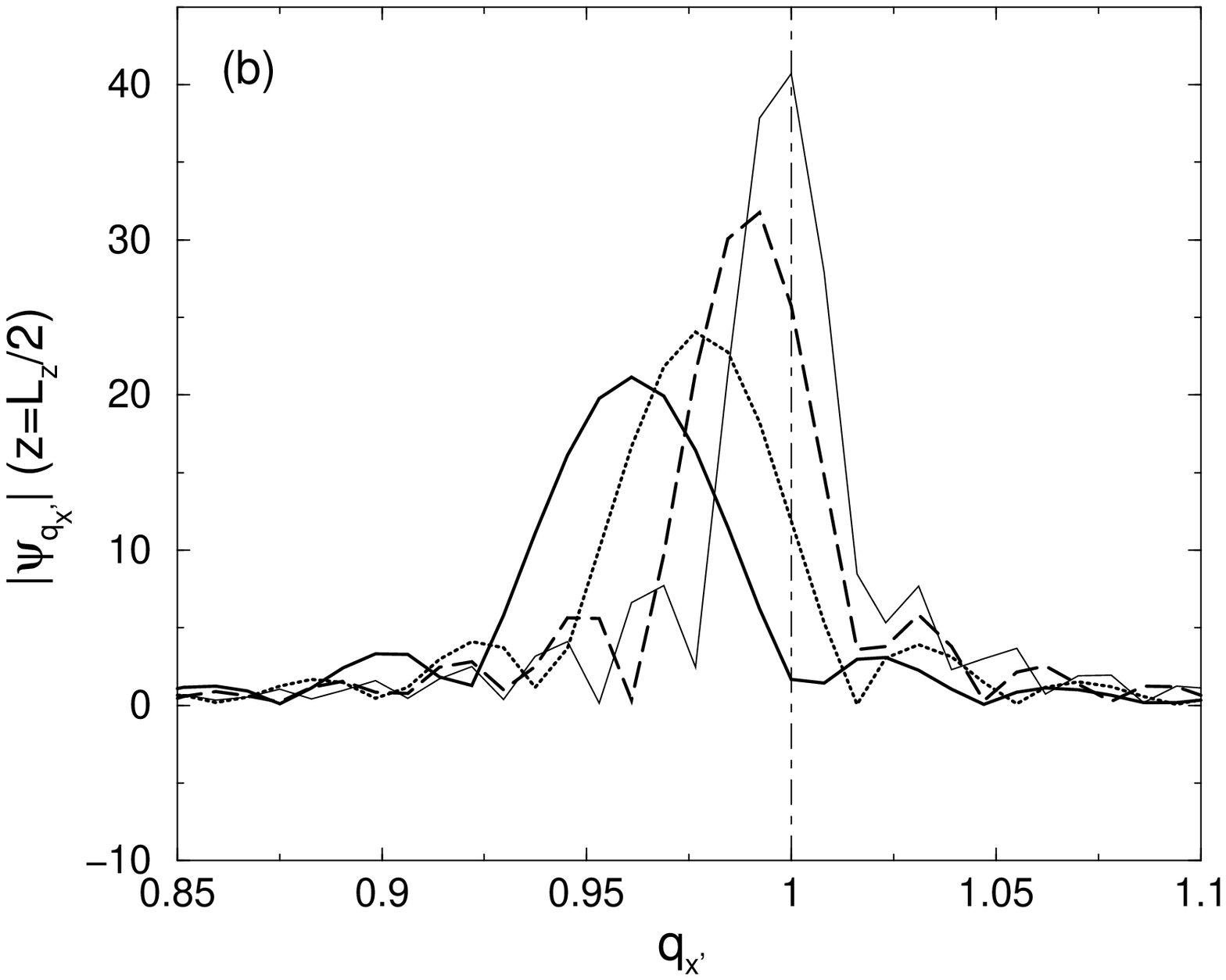,width=4.in}}
\caption{One dimensional structure factor for transverse lamellae 
$|\psi_{q_{x'}}|$ along the
$x'$ direction at $z'=L_{z'}/2$ as a function of wavenumber $q_{x'}$. 
(a), different times from 
top to bottom: $t=30 T_0$, $40 T_0$, $50 T_0$ and $60 T_0$, 
with $\gamma=0.4$ and $\omega=0.04$.
(b), different strain amplitudes: $\gamma=0.4$ and $t=100 T_0$ (solid
curve), $\gamma = 0.3$ and $t=300 T_0$ (dotted curve),
$\gamma = 0.2$ and $t=1200 T_0$ (dashed curve), and $\gamma = 0.1$ and
$t=12000 T_0$ (thin solid curve). Here $\omega=0.1$, and 
$\Delta t=0.1$ except for $\gamma=0.1$ in which case we have used
$\Delta t=0.2$. 
In both cases (a) and (b), the vertical dot-dashed line indicates the location of 
the wavenumber $q_{x'}=q_0=1$. 
}
\label{fig_qx}
\end{figure}

\begin{figure}
\centerline{\epsfig{figure=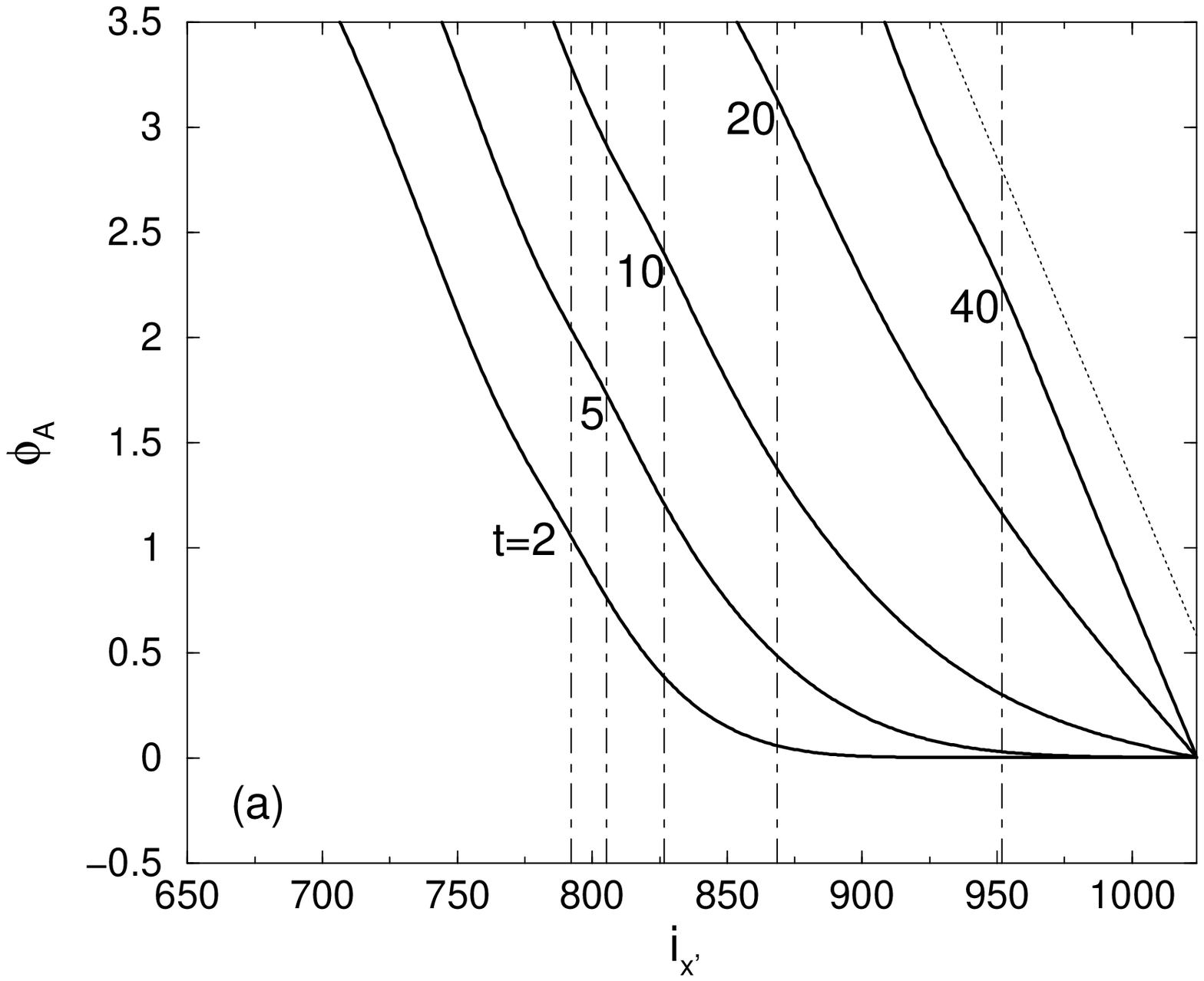,width=4.in}}
\centerline{\epsfig{figure=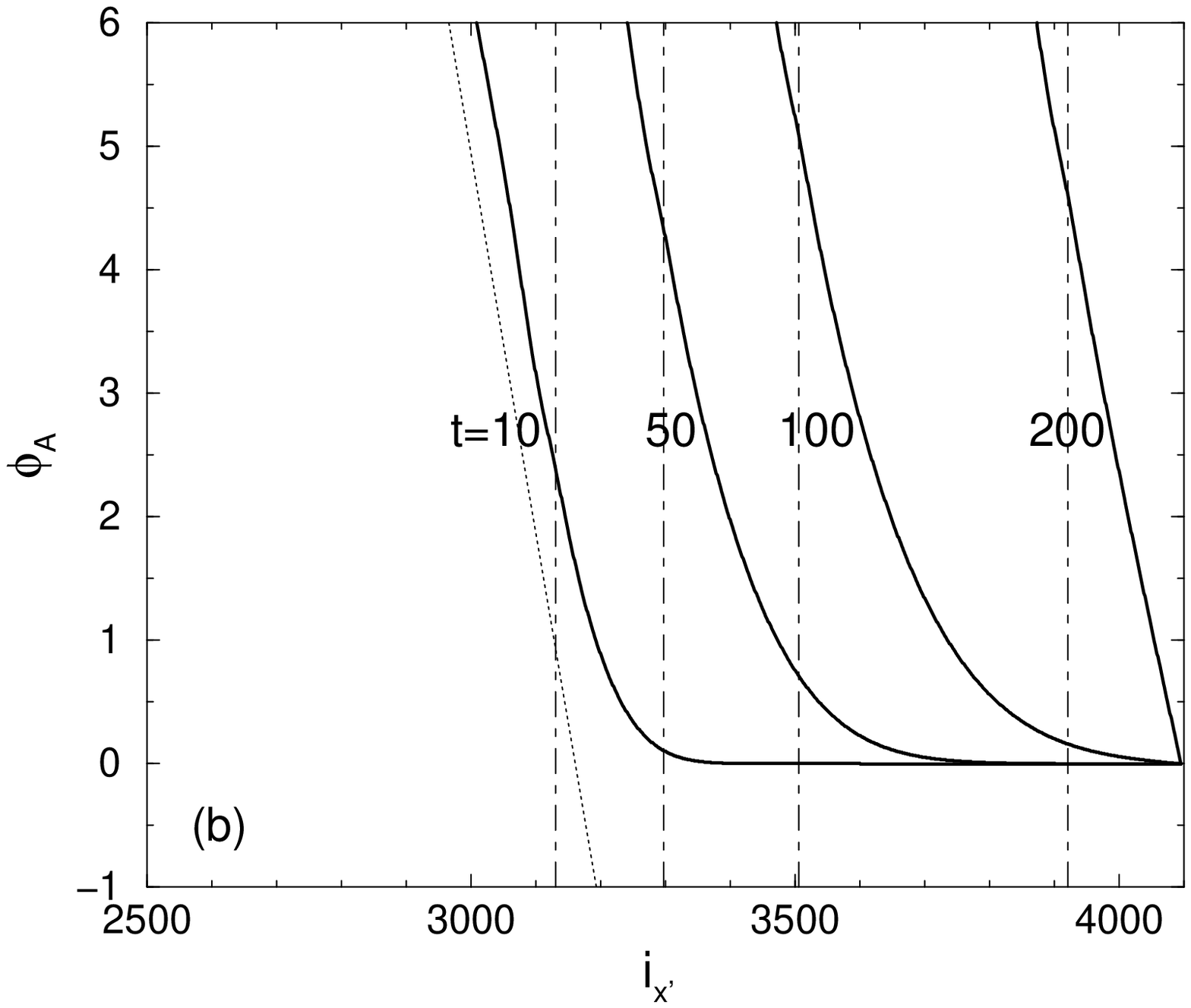,width=4.in}}
\caption{Phase $\phi_A$ of the complex amplitude $A$ as a function of
position in the sheared frame (in terms of the grid index $i_{x'}$)
with $\Delta x'= \lambda_0/8$, $\gamma=0.4$, and $\omega=0.04$. 
Two system sizes are shown: (a) $L=1024$ at times 
(from left to right): $t=2 T_0$, $5 T_0$, $10 T_0$, $20 T_0$, and 
$40 T_0$ (to be compared with Fig. \ref{fig_qx}(a)); and (b) $L=4096$
at times (from left to right) $t=10 T_0$, $50 T_0$, $100 T_0$, and 
$200 T_0$. The vertical dot-dashed lines indicate the instantaneous grain boundary 
positions $x'_{\rm gb}$, and a dotted line with slope $-\delta
q^m_{x'} \cdot \Delta x'=-5\Delta x'/128$ is also shown for reference.
}
\label{fig_phaseA}
\end{figure}

As the boundary moves over time, we have observed that the wavenumber
of the transverse lamellae does not remain constant and equal to
$q_0$. Instead, it is slowly readjusted by wavenumber diffusion, as
shown by both direct solution of the coarse grained model 
of Sec. \ref{sec_num} and the corresponding complex amplitude 
equations of Sec. \ref{sec_ampl}. The wavenumber shift 
$\delta q^m_{x'}$ is approximately independent of shear frequency $\omega$,
but strongly dependent on the strain amplitude $\gamma$. In order to understand
the physical origin of wavenumber compression (or 
lamellae expansion) that occurs in the transverse region A, we
focus on the lamellae near the grain boundary, since the calculation 
in Sec. \ref{sec_ampl} shows that
wavenumber change is initiated at the boundary, and then propagates into 
the bulk. Since the amplitude of the transverse lamellae go to zero 
at the boundary region, it is possible to create or eliminate lamellar
planes there in a way that it is not possible in the bulk for the parameters of
our study \cite{drolet99}. First consider the stability diagram of the
Swift-Hohenberg model (\ref{eq_s-h}) at zero shear
\cite{greenside84}. For fixed $\epsilon$ and $q_x > q_0$, the closest
instability boundary in the $q_x$-$\epsilon$ diagram is the Eckhaus
instability given by
\begin{equation}
\epsilon = 12 (q_x - q_0)^2.
\label{eq_Eck}
\end{equation}
For $\epsilon=0.04$ and $q_0=1$, we have the Eckhaus boundary at
$q_x^{\rm E} = q_0+(\epsilon /12)^{1/2}
=1.0577$. In our case $q_x$ is the wavenumber of the transverse 
lamellae in the lab frame, and equal to $q_0 \sqrt{1+a^2(t)}$ if we
assume that the lamellae are rigidly distorted by the shear.
The maximum value of $q_x$ is then $q_x^{\rm
max}=q_0\sqrt{1+\gamma^2}$ so that 
for $\gamma=0.4$ we have $q_x^{\rm max} > q_x^{\rm E}$. Although this
is not sufficient to destabilize the bulk transverse region (according
to the stability diagrams in Ref. \onlinecite{drolet99} obtained by
Floquet analysis over the entire period of the oscillation), it
appears to be sufficient to induce lamellar elimination at the grain
boundary region, as seen in Figs. \ref{fig_qx} and
\ref{fig_phaseA}. The figures show that $\delta q^m_{x'}=5 \Delta q$,
corresponding to the elimination of $5$ transverse lamellae. As $\gamma$ 
decreases $q_x^{\rm max}$ also decreases, becoming smaller
than $q_x^{\rm E}$, and eventually lamellae elimination is expected to cease.
This is consistent with our results shown in Fig. \ref{fig_qx}(b). With
decreasing value of $\gamma$ from $0.4$ to $0.1$, the number of 
lost transverse lamellae decreases from $5$ to $0$.

In summary, for small shear strains and low frequencies such that 
diffusion of order parameter is of the same order as advection by the
flow, the excess free energy in the transverse region relative to the
marginal parallel region is dissipated through order parameter
diffusion in the grain boundary region. The latter includes break up
and reconnection of transverse lamellae, and a weak Eckhaus instability 
developing at the grain boundary that diffuses into the bulk
transverse lamellae leading to dynamical wavenumber readjustment. A
weakly nonlinear analysis, as well as the amplitude equations derived, 
capture quantitatively all aspects of grain
boundary motion, including the boundary velocity and the wavenumber
readjustment. The order parameter distribution in the boundary region
can be represented crudely by introducing an
adiabatic approximation into the amplitude equations, which gives a
reasonable approximation to the net boundary velocity toward the 
transverse region, and be well reproduced by the direct solution
of the amplitude equations.
Although our study is confined to the case of a transverse/parallel
grain boundary in two dimensions, we expect that our results will
qualitatively hold for both three dimensional transverse/parallel 
and transverse/perpendicular cases. In three dimensions, however, 
there is a completely different type of tilt boundary, that between
parallel and perpendicular lamellae. Both orientations are marginal with
respect to the shear. This configuration is currently under 
investigation.

\begin{acknowledgments}
This work was supported by the National Science Foundation 
under grant DMR-0100903.
\end{acknowledgments}

\newpage

\end{document}